\documentclass[aps,prl,floatfix,twocolumn,showpacs,reprint]{revtex4-1}
\usepackage{amsfonts}
\usepackage{mathrsfs}
\usepackage{amsmath}% needed for subequations
\usepackage{color}
\usepackage{graphicx}
\usepackage{bm}% bold maths
\usepackage{amssymb}
\usepackage{xspace}
\usepackage{epstopdf}
\usepackage{dcolumn}% Align table columns on decimal point
\usepackage{longtable}
\usepackage{multirow}
\usepackage[colorlinks=true, letterpaper=true, pdfstartview=FitV, linkcolor=blue, citecolor=blue, urlcolor=blue]{hyperref}

\makeatletter

\newcommand{\Rmnum}[1]{\expandafter\@slowromancap\romannumeral #1@}
\makeatother
\begin{document}
\title{
  Anomalous Topological Pumps and Fractional Josephson Effects
       }
\author{Fan Zhang}%\email{zhf@sas.upenn.edu}
\author{C. L. Kane}
\affiliation{Department of Physics and Astronomy, University of Pennsylvania, Philadelphia, PA 19104, USA}
\begin{abstract}
We discover novel topological pumps in the Josephson effects for superconductors.
The phase difference, which is odd under the chiral symmetry defined by the product of time-reversal and particle-hole symmetries, acts as an anomalous adiabatic parameter.
These pumping cycles are different from those in the ``periodic table", and are characterized by $\mathbb{Z}\times\mathbb{Z}$ or $\mathbb{Z}_2\times\mathbb{Z}_2$ strong invariants.
We determine the general classifications in class AIII, and those in class DIII with a single anomalous parameter.
For the $\mathbb{Z}_2\times\mathbb{Z}_2$ topological pump in class DIII,
one $\mathbb{Z}_2$ invariant describes the coincidence of fermion parity and spin pumps
whereas the other one reflects the non-Abelian statistics of Majorana Kramers pairs,
leading to three distinct fractional Josephson effects.
\end{abstract}
%\date{\today}
%\pacs{03.65.Vf, 03.75.Lm, 02.40.Pc, 03.67.Lx}
% 03.65.Vf, Phases: geometric; dynamic or topological
% 03.75.Lm Josephson effect, BEC, solitons, vortices, and topological excitations
% 02.40.Pc General topology
% 03.67.Lx Quantum computation architectures and implementations
\maketitle
{\color{cyan}\indent{\em Introduction.}}---
Topological insulators (TI) and superconductors (SC) have attracted tremendous interest~\cite{RMP1,Nature,RMP2} in condensed matter physics.
Gapped electronic Hamiltonians subject to time-reversal (TR) and/or particle-hole (PH) symmetries can be classified topologically.
Nontrivial topological classes are associated with protected gapless modes on the boundary.
The topological phases for gapped free fermion systems with different symmetries and dimensions fit together into
an elegant ``periodic table"\cite{Ryu,Kitaev,Teo} that unifies and generalizes the integer quantum Hall states~\cite{CN1,CN2},
the chiral $p$ wave SCs~\cite{Read,Kitaev-2001,Ivanov}, and the $\mathbb{Z}_2$ TIs~\cite{Z2}.
This framework has been extended to classify topological defects and pumping cycles~\cite{Teo},
which are characterized by a Hamiltonian $\mathcal{H}({\bm k},{\bm r})$.
Here ${\bm k}$ is a $d_k$ dimensional momentum variable defined in the Brillouin zone (BZ),
whereas ${\bm r}$ is a set of $d_r$ adiabatic parameters describing spatial and/or temporal variation of the Hamiltonian.
${\bm k}$ and ${\bm r}$ are distinguished by their behaviors under TR and PH symmetries,
i.e., ${\bm k}\rightarrow-{\bm k}$ whereas ${\bm r}\rightarrow{\bm r}$.
It was found that the topological classes for $\mathcal{H}({\bm k},{\bm r})$ only depends on the combination $d_k-d_r$~\cite{Teo}.
Thus, the invariants characterizing defects and pumps for $d_r \ne 0$ are related to the invariants
(given by $\mathbb{Z}$, $\mathbb{Z}_2$, or $0$) in the original table with $d_r=0$.

In this Letter we introduce a class of adiabatic pumping cycles with anomalous parameters that have a mixed behavior under TR and PH symmetries.
Such a pump naturally arises in the Josephson effect for SCs that respect TR symmetry (i.e. class DIII).
Consider a Josephson junction in which the phase difference $\phi$ is an adiabatic parameter.
Since $\phi$ is odd under TR, a $2\pi$ cycle crosses two TR invariant points at $\phi=0$ and $\pi$,
similar to the $\mathbb{Z}_2$ spin pump~\cite{spin-pump}.
However, unlike the spin pump, the Bogoliubov de Gennes Hamiltonian has PH symmetry for each $\phi$, so $\phi$ is even under PH.
Unlike both $k$ and $r$, $\phi$ is odd under the combination of TR and PH, which defines the unitary chiral symmetry.
We will refer parameters with this property as anomalous.
One may also consider another type of anomalous parameter $\theta$ which is even (odd) under TR (PH),
and anticipate a new table which should depend on $d_k-d_r$ and $d_{\phi}-d_{\theta}$.

\begin{figure}[t!]
\scalebox{0.56}{\includegraphics*{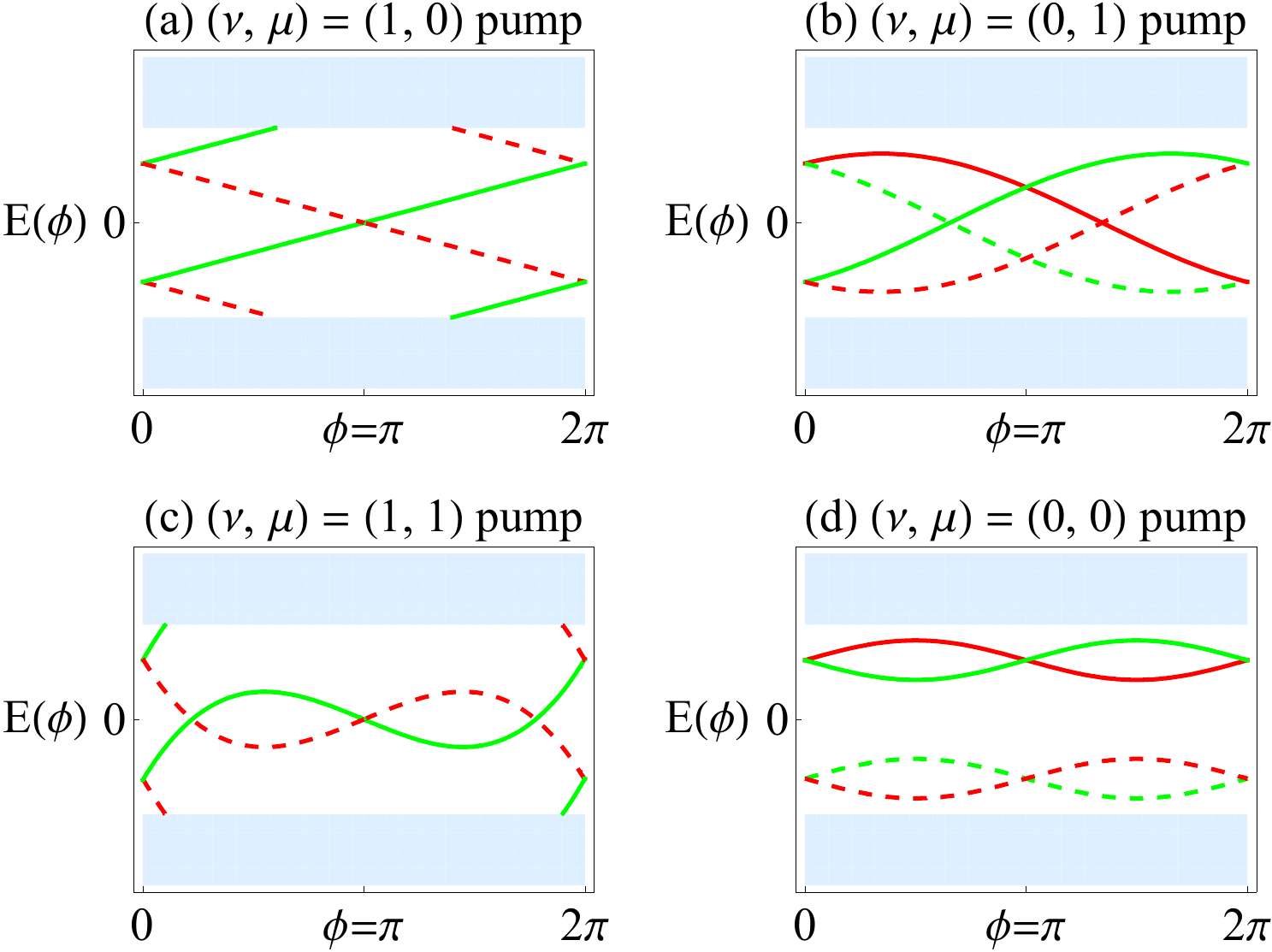}}
\caption{The three $\mathbb{Z}_2$ topological and one trivial pumps in the $\mathbb{Z}_2\times\mathbb{Z}_2$ Josephson effects for class DIII SCs.
Green and red (solid and dashed) states are related by the TR (PH) symmetry. The shaded areas are the bulk continuum.}
\label{fig1}
\end{figure}

We find that anomalous parameters lead to topological classes that substantially differ from those in the original table.
We work out the general classification in class AIII (which only has the chiral symmetry)
and show that the classification is $\mathbb{Z}\times\mathbb{Z}$ when the numbers of normal and anomalous parameters are both odd.
We further determine the topological pumps in class DIII, with a single anomalous parameter $\phi$.
In particular, for class DIII with $d_k=d_\phi=1$ we show there is a $\mathbb{Z}_2\times \mathbb{Z}_2$ strong topological invariant.
One $\mathbb{Z}_2$ invariant describes a TR (PH) invariant version of the fermion parity (spin) pump
whereas the other $Z_2$ reflects the non-Abelian statistics of Majorana Kramers pairs,
leading to three distinct $4\pi$ periodic Josephson effects, as TR invariant topological pumping cycles in SCs.
Our main results are summarized in Fig.~\ref{fig1} and Tables~\ref{tableone}~and~\ref{tabletwo}.

{\color{cyan}\indent{\em $\mathbb{Z}\times\mathbb{Z}$ invariant.}}---
We first analyze the simplest case, class AIII, in which antiunitary symmetries are absent and show the chiral symmetry ($\Pi$) leads to a $\mathbb{Z}\times\mathbb{Z}$ invariant.
We will use this result later to derive a $\mathbb{Z}_2\times\mathbb{Z}_2$ invariant in class DIII.
Moreover, on its own it can be used to classify the pumps in Josephson effects with a mirror symmetry, in which
each mirror eigenspace by itself has chiral symmetry, but no TR or PH symmetry.

Consider a gapped Hamiltonian $\mathcal{H}(k,\phi)$ that satisfies
\begin{eqnarray}
\Pi^{-1}\mathcal{H}(k,\phi)\Pi&=&-\mathcal{H}(k,s\phi)\,,\label{CS}
\end{eqnarray}
where $s=\pm$.
Focusing on the strong invariant, we may think of $k$~($\phi$) as the azimuth (polar) angle of a sphere.
The chiral symmetry (\ref{CS}) requires
the valence (v) band and conduction (c) band Berry curvature to satisfy $\mathcal{F}^v(k,\phi)=s\mathcal{F}^c(k,s\phi)$,
whereas the completeness relation of wave functions restricts $\sum_{i=c,v}\mathcal{F}^i(k,\phi)=0$,
leading to $\mathcal{F}^v(k,\phi)=-s\mathcal{F}^v(k,s\phi)$.
Consequently, for normal cases in which $s=+$, i.e., in the original table with $d=2$~\cite{Ryu,Kitaev,Teo}, the Chern number must vanish,
whereas for anomalous parameters with $s=-$ the Chern number survives.
On the other hand, the equator $\mathcal{H}(k,0)$ describes a normal one-dimensional insulator in class AIII,
which has an integer winding number~\cite{Ryu,Volovik}.

Now we demonstrate that the Chern number $N_c$ and the winding number $N_w$ are distinct but related.
Winding numbers are well known to be gauge dependent.
However, we can choose a gauge in which the wave functions of the equator are able to contract to a nonsingular north pole.
This can be accomplished~\cite{TMSC} by introducing a continuous deformation that trivializes $\mathcal{H}(k,0)$ to a constant $\Pi$,
formulated by $\mathcal{H}_0(k,\phi')=\mathcal{H}(k,0)\cos\phi'+\Pi\sin\phi'$.
In this trivial gauge, Stokes' theorem relates the loop integral of Berry connection along the equator to
the integral of Berry curvature over the upper hemisphere (u.h.)
\begin{eqnarray}
\pi N_w=\oint\mathcal{A}^v(k)dk=\int_{u.h.}\mathcal{F}_0^v(k,\phi')dkd\phi'\,.\label{I2}
\end{eqnarray}
Moreover, a Chern number is produced if one glues together the two u.h. integrals of $\mathcal{F}_0^v(k,\phi')$
and $\mathcal{F}^v(k,\phi)$ along the common equator, i.e.,
\begin{eqnarray}
2\pi N_d=\int_{u.h.}\mathcal{F}_0^v(k,\phi')dkd\phi'-\int_{u.h.}\mathcal{F}^v(k,\phi)dkd\phi\,.\label{I1}
\end{eqnarray}
As a result of $\mathcal{F}^v(k,\phi)=\mathcal{F}^v(k,-\phi)$ derived above, the last integral is $\pi N_c$.
We therefore conclude
\begin{eqnarray}
N_w=N_c+2N_d\,.\label{WC}
\end{eqnarray}
Thus, the winding number along the equator and the Chern number over the sphere are distinct but only differ by an even integer.
To further illustrate this unprecedented relation,
consider the following smooth and nonsingular Hamiltonian flattened on a unit sphere
\begin{equation}
\mathcal{H}=\cos(3\phi)[\cos(Nk)\sigma_x + \sin(Nk)\sigma_y] + \sin(3\phi)\sigma_z\,.\label{Hnm}
\end{equation}
Here $\Pi=\sigma_z$ and the integer $N(\phi)$ is $N_w$ when $|\phi|\leq\pi/6$ and switches to $N_d$ otherwise.
Along the equator $\mathcal{H}(k,0)$ has a winding number $N_w$ in the trivial gauge,
whereas over the sphere $\mathcal{H}(k,\phi)$ has a Chern number $N_c=N_w-2N_d$.
Note that $(N_w\pm N_c)/2$ are independent integers.

This result can be generalized to higher ``dimensions".
When $d_k+d_{\phi}=2n$ is even, Eq.~(\ref{CS}) with $s=-$ leads to
\begin{eqnarray}
{\rm Tr}[\mathcal{F}({\bm k},{\bm \phi})^n]=(-1)^{d_{\phi}+1}\,{\rm Tr}[\mathcal{F}({\bm k},-{\bm \phi})^n]\,,\label{F}
\end{eqnarray}
where the Berry curvature $\mathcal{F}$ is $d\mathcal{A}+\mathcal{A}\wedge\mathcal{A}$
and $\mathcal{A}$ is the non-Abelian Berry connection.
The $n$-th Chern number is proportional to the integral of Eq.~(\ref{F})
over the extended BZ spanned by $\bm k$ and $\bm\phi$.
Thus, the Chern number vanishes when $d_{\phi}$ is even,
whereas it survives when $d_{\phi}$ is odd.
When $d_k$ is odd, there is a winding number along the equator spanned by $\bm k$.
These two invariants follow from the homotopy groups $\pi_{d_k}$ and $\pi_{d_k+d_{\phi}}$
of the valence band structure, i.e., a map from the extended BZ to the space of $\mathcal{H}({\bm k},{\bm \phi})$.
These results, summarized in Table~\ref{tableone}, exhibit a periodicity $2$ in both $d_{\phi}$ and $d_k$.
\begin{table}[b!]
\caption{Topological classifications of gapped Hamiltonians in class AIII with and without anomalous parameters.}
% Adds a space between the text and the [T]op \hline
\newcommand\T{\rule{0pt}{3.0ex}}
% Adds a space between the text and the [B]ottom \hline
\newcommand\B{\rule[-1.7ex]{0pt}{0pt}}
\centering
%\begin{ruledtabular}
\begin{tabular}{c|c|cc}
      \hline\hline
      \multicolumn{2}{c|}{$d_k-d_r$} & \quad even \quad & \quad odd \quad \T\\[1pt]
      \hline
      \multirow{2}{*}{$d_{\phi}-d_{\theta}$\T} & \; even \; & \quad $0$ \quad & \quad $\mathbb{Z}$ \T\\
      %\cline{2-4}
      & \; odd \; & \quad $0$ \quad & \quad $\mathbb{Z}\times\mathbb{Z}$ \T\\[1pt]
      \hline\hline
\end{tabular}
%\end{ruledtabular}
\label{tableone}
\end{table}

{\color{cyan}\indent{\em $\mathbb{Z}_2\times\mathbb{Z}_2$ Josephson effect.}}---
Before advancing the $\mathbb{Z}_2\times\mathbb{Z}_2$ strong invariant for class DIII with $d_{\phi}=d_k=1$,
we first analysis the boundary consequence, i.e., the spectra of Andreev bound states (ABS) as a function of the phase difference $\phi$,
which naturally arise from the Josephson effects for topological SCs with TR symmetry~\cite{TRITSC,WTI,Wong,Nagaosa,Berg}.

Fig.~\ref{fig1}(a) describes the spectrum of two ABSs with a single crossing at $\phi=\pi$ and $E=0$.
This twist reminds us of the $4\pi$ periodic Josephson effect~\cite{Kitaev-2001,Kane-D,Lutchyn-D,Oreg-D},
yet the switching must occur at $\phi=\pi$ (or $0$) required by the TR symmetry.
This crossing is also reminiscent of the edge state of quantum spin Hall insulator~\cite{RMP1,Nature,RMP2,Z2},
with an extra feature being the PH symmetry of the conduction and valence bands.
This pump is even robust against TR (PH) symmetry breaking, as long as one symmetry is intact.
Thus this pump is the coincidence of spin and fermion parity pumps.
A symmetry-allowed perturbation can only gap even number of such spectra,
indicating that this pumping cycle is characterized by a $\mathbb{Z}_2$ index $\nu=1$.
Such a topological pump can be realized by proximity coupling TI edge states~\cite{Kane-D}
or hybridized double Rashba wires~\cite{Berg} to a $s$ wave Josephson junction.

Fig.~\ref{fig1}(b) depicts the spectrum of four ABSs exhibiting a pair of zero-energy crossings.
The degeneracies at $\phi=0,~\pi$ are required by the TR symmetry,
whereas the crossings at $E=0$ are protected by local conservation of fermion parity.
By examining a model below, we find that the zero-energy crossings can not be annihilated
without breaking the TR or PH symmetry, even if they are brought together.
However, even number of such double-crossings can be gapped out by a symmetry-allowed disturbance.
These features, in sharp contrast to those in Fig.~\ref{fig1}(a), imply a distinct $\mathbb{Z}_2$ index $\mu=1$.
Importantly, this topological pump explicitly shows the non-Abelian statistics of Majorana Kramers pairs protected by the TR symmetry.
In the fermion parity ($0$ or $1$) basis of each Kramers partner
($\uparrow$ or $\downarrow$), the adiabatic pumping of fermion parity and spin follows
\begin{eqnarray}
|0_{\uparrow}0_{\downarrow}\rangle\rightarrow|1_{\uparrow}0_{\downarrow}\rangle\rightarrow
|1_{\uparrow}1_{\downarrow}\rangle\rightarrow|1_{\uparrow}0_{\downarrow}\rangle\rightarrow|0_{\uparrow}0_{\downarrow}\rangle\,,
\end{eqnarray}
in which $\phi$ advances by $\pi$ in each step.
This topological pump can be achieved~\cite{TRITSC,breaking-mirror} through proximity coupling a Rashba wire to a $s_{\pm}$ wave Josephson junction.

Since $\mu$ and $\nu$ are independent invariants,
it is possible to have a third $\mathbb{Z}_2$ topological pump with $\nu=\mu=1$, as shown in Fig.~\ref{fig1}(c).
For comparison, the trivial pump is plotted in Fig.~\ref{fig1}(d).
Although when $\nu=1$ the ABSs inevitably couple to the bulk continuum in a pumping cycle,
the $\mathbb{Z}_2\times\mathbb{Z}_2$ Josephson effect, particularly the non-Abelian statistics of Majorana Kramers pairs,
will lead to intriguing braiding~\cite{Ivanov,Nayak,Fisher} of Majoranas when the TR symmetry is restored at special points.

\begin{figure}[t!]
\scalebox{0.40}{\includegraphics*{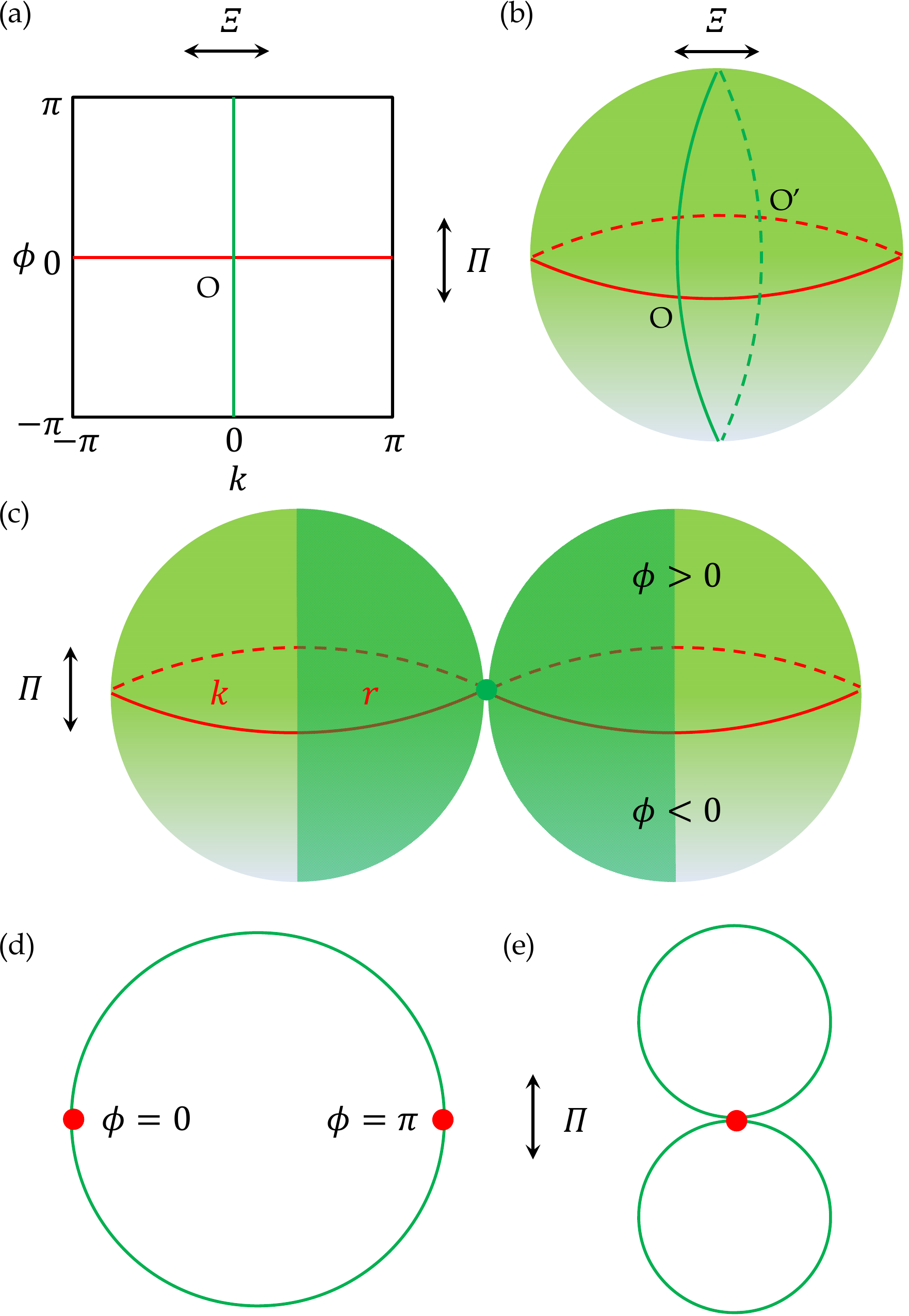}}
\caption{(a) The extended BZ of $\mathcal{H}(k,\phi)$ in class DIII.
The torus (a) becomes a sphere (b) if the $k,\phi=\pi$ lines are trivialized and contract to a point $O'$.
If the trivial $k=0$ circle further contracts to a point,
the sphere (b) continuously deforms into (c) two spheres related by the PH symmetry.
In each sphere there is a $\mathbb{Z}\times\mathbb{Z}$ invariant,
whereas the deformation from (b) to (c) has a $2\mathbb{Z}\times2\mathbb{Z}$ ambiguity.
Hence $\mathcal{H}(k,\phi)$ is characterized by a $\mathbb{Z}_2\times\mathbb{Z}_2$ strong invariant.
If the trivial $\phi=0$ and $\pi$ points are glued together,
the $k=0$ circle (d) becomes the two circles in (e) with each being trivial,
proving the above statement that the $k=0$ circle is trivial.}
\label{fig2}
\end{figure}

{\color{cyan}\indent{\em Homotopy argument.}}---
Now we derive the above $\mathbb{Z}_2\times\mathbb{Z}_2$ strong invariant using a homotopy argument in the spirit of
the Moore-Balents~\cite{Moore-Balents} argument on the $\mathbb{Z}_2$ TIs.
When folded into each other, the two SCs coupled at a Josephson junction may be described by a Hamiltonian $\mathcal{H}(k,\phi)$ in class DIII.
The Josephson effects sketched in Fig.~\ref{fig1} can thus be interpreted as the boundary consequences of the bulk invariant of
$\mathcal{H}(k,\phi)$, which inherits PH and TR symmetry constraints as follows:
\begin{eqnarray}
\Xi^{-1}\mathcal{H}(k,\phi)\Xi&=&-\mathcal{H}(-k,\phi)\,,\label{PHS}\\
\Theta^{-1}\mathcal{H}(k,\phi)\Theta&=&\mathcal{H}(-k,-\phi)\,,\label{TRS}
\end{eqnarray}
the combination of which determines the chiral symmetry specified by Eq.~(\ref{CS}) with $s=-$.
We focus on deriving the strong invariants instead of the weak ones, so
we assume the topologically nontrivial physics only occurs near $k,\phi=0$.
The $k,\phi=\pi$ lines may be trivialized and contract to a point $O'$.
Hence the torus in Fig.~\ref{fig2}(a), i.e., the extended BZ of $\mathcal{H}(k,\phi)$, is topologically equivalent to a sphere in Fig.~\ref{fig2}(b).
Because of the topological triviality of $\mathcal{H}(0,\phi)$ which will be demonstrated below,
the $k=0$ circle may further contract to a trivial point, as described in Fig.~\ref{fig2}(c).
The resulting two spheres are related by the PH (TR) symmetry,
whereas for each one only the chiral symmetry is unbroken.
As we have demonstrated, there is a $\mathbb{Z}\times\mathbb{Z}$ strong invariant for each sphere.

\begin{table}[b!]
\caption{Topological classifications of adiabatic pumps in class DIII, with zero or one anomalous parameter effectively.}
% Adds a space between the text and the [T]op \hline
\newcommand\T{\rule{0pt}{3.0ex}}
% Adds a space between the text and the [B]ottom \hline
\newcommand\B{\rule[-1.7ex]{0pt}{0pt}}
\centering
%\begin{ruledtabular}
\begin{tabular}{c|c|ccccc}
      \hline\hline
      \multicolumn{2}{c|}{$(d_k-d_r)$ mod $8$} & $0,4,5,6$ & $1$ & $2$ & $3$ & $7$\T\\[1pt]
      \hline
      \multirow{2}{*}{$d_{\phi}-d_{\theta}$} & 0 & $0$ & $\mathbb{Z}_2$ & $\mathbb{Z}_2$ & $\mathbb{Z}$ & $2\mathbb{Z}$\T\\
      %\hline
       & $1$ & $0$ & $\mathbb{Z}_2\times\mathbb{Z}_2$ & $\mathbb{Z}_2\times\mathbb{Z}_2$ & $\mathbb{Z}\times\mathbb{Z}$ & $2\mathbb{Z}\times2\mathbb{Z}$\T\\[1pt]
      \hline\hline
\end{tabular}
%\end{ruledtabular}
\label{tabletwo}
\end{table}

However, there are multiple topologically inequivalent contractions from the $k=0$ circle to a point.
These ambiguities reduce the $\mathbb{Z}\times\mathbb{Z}$ invariant to $\mathbb{Z}_2\times\mathbb{Z}_2$ invariant.
For a homotopic deformation, it is required that at each stage the contracted circle has the same symmetry constraints as the original one.
Thus the contraction is naturally parameterized by $r$ which is even under both TR and PH,
as shown in Fig.~\ref{fig2}(c). The two hemispheres forming in the contraction can be glued into a sphere,
and the corresponding Hamiltonian is $\mathcal{H}(r,\phi)$ in class DIII.
From Table~\ref{tableone}, $\mathcal{H}(r,\phi)$ has a $\mathbb{Z}\times\mathbb{Z}$ invariant.
Yet the antiunitary symmetries require both integers to be even.
Indeed the original table~\cite{Teo} has revealed that the winding number of $\mathcal{H}(r,0)$ in class DIII is $2\mathbb{Z}$,
and that the Chern number of $\mathcal{H}(r,\phi)$ is also $2\mathbb{Z}$ in both class D and AII.
Similarly to above, here the winding number and the Chern number are distinct but only differ by $4\mathbb{Z}$.
This not only explains there are $2\mathbb{Z}\times2\mathbb{Z}$ topologically distinct contractions of the $k=0$ circle,
but also determines the invariant of $\mathcal{H}(r,\phi)$ in class DIII.

The remaining task is to show $\mathcal{H}(\phi)$ is topologically trivial in class DIII.
From the original table~\cite{Ryu,Kitaev,Teo}, zero-dimensional DIII SCs are trivial
and thus we can glue the two points $\phi=0$ and $\pi$ together,
as done in Fig.~\ref{fig2}(e). The two resulting circles are related by TR (chiral) symmetry,
whereas each one has PH symmetry, i.e., in class D.
As $\mathcal{H}(\phi)$ in class D is trivial shown in the table~\cite{Teo},
we conclude that $\mathcal{H}(\phi)$ is indeed trivial in class DIII.

So far, we have established the strong invariants for $\mathcal{H}(k,\phi)$,
$\mathcal{H}(r,\phi)$, and $\mathcal{H}(\phi)$ in class DIII, which are summarized in Table~\ref{tabletwo}.
Now we demonstrate the remaining two nontrivial entries in this new Table.
Consider the case for $d_\phi=1$ and $d_k=3$ in Table~\ref{tableone},
the second Chern number is compatible with having TR and PH symmetries,
as this invariant exists in class AII with $d_k+d_{\phi}=4$~\cite{Kitaev,reduction} and in class D with $d_k-d_{\phi}=2$~\cite{Teo}.
The winding number along the equator is exactly the integer invariant in class DIII with $d_k=3$~\cite{Ryu,Volovik}.
Hence $\mathcal{H}({\bm k},\phi)$ with $d_k=3$ in class DIII also has a $\mathbb{Z}\times\mathbb{Z}$ invariant.
As for $\mathcal{H}({\bm k},\phi)$ with $d_k=2$ in class DIII,
there can be a weak $\mathbb{Z}_2\times\mathbb{Z}_2$ invariant in each of the two ``planes" with $d_k=d_{\phi}=1$,
as we have demonstrated above. If one plane is trivial then the invariant in the other plane becomes a strong invariant,
which is analogous to the relation between two- and three-dimensional $\mathbb{Z}_2$ TIs~\cite{Moore-Balents}.
In light of these analyses, we complete Table~\ref{tabletwo} which has Bott periodicity $8$ in $d_k-d_r$~\cite{Kitaev,Teo}.

{\color{cyan}\indent{\em Effective theory.}}---
Table~\ref{tabletwo} suggests a dimension reduction rule, generalizing the case~\cite{reduction} with no anomalous parameter.
This becomes more clear in a minimal effective theory near ${\bm k}=0$ and $\phi=\pi$,
which is built as linear combinations of anticommuting generators.
This Clifford algebra allows us to relate the strong invariant to the winding degree in maps between spheres.
We choose a gauge in which the PH and TR operators are $\Xi=\mathcal{K}$ and $\Theta=\sigma_y\mathcal{K}$.
Consider a four (eight) band model
\begin{equation}
\mathcal{H}=k_x\sigma_x s_x+k_z\sigma_z+\delta\phi\sigma_y(\tau_z)
+k_y\sigma_x s_z+M\sigma_x s_y\,,\label{model}
\end{equation}
where $M=m-{\bm k}^2-\delta\phi^2$ and $\hat{k}_y$ is normal to the Josephson junction.
The boundary problem is specified by the last two terms in Eq.~(\ref{model}), with $m$ switching signs~\cite{boundary}.

The four-band model has $N_c=N_w=1$, and we can derive the spectrum of ABSs near $\phi=\pi$
\begin{eqnarray}
\bar{\mathcal{H}}=k_x\sigma_x+k_z\sigma_z+\delta\phi\sigma_y\,,\label{2band}
\end{eqnarray}
which resembles a ``Weyl fermion".
Any perturbation in $\bar{\mathcal{H}}$ is prohibited by {\it both} TR and PH symmetries.
Even even number of such spectra can not be gapped, consistent with the invariants being integers.
When one or two $k$ terms are taken off, $\bar{\mathcal{H}}$ describes the lower dimensional cases.
Although any disturbance is still prohibited, a pair of such spectra can be gapped without breaking any symmetry,
indicating the $\mathbb{Z}_2$ character.
The $d_k=1$ case exactly describes the topological pump in Fig.~\ref{fig1}(a).

The eight-band model has $N_c=0$ and $N_w=2$, and the corresponding ABSs can be described by
\begin{eqnarray}
\widetilde{\mathcal{H}}=k_x\sigma_x+k_z\sigma_z+\delta\phi\sigma_y\tau_z\,,\label{4band}
\end{eqnarray}
resembling a ``Dirac fermion". In the presence of {\it both} TR and PH symmetries,
a perturbation in $\widetilde{\mathcal{H}}$ is allowed but can not gap the spectrum for the $d_k=1,2,3$ cases.
Without breaking a symmetry, two copies of $\widetilde{\mathcal{H}}$ can be gapped for the $d_k=1,2$ cases but not for the $d_k=3$ case,
reflecting their $\mathbb{Z}_2$ and integer invariants, respectively.
For the $d_k=1$ case, the symmetry-allowed couplings $\Delta\sigma_{x,z}\tau_y$ only split the Majorana quartet at $\delta\phi=0$
into a pair of zero-energy crossings, as shown in Fig.~\ref{fig1}(b).

Evidently, in the $d_k=3$ cases, the protection of the gapless nature of $\bar{\mathcal{H}}$ requires no symmetry,
whereas that of $\widetilde{\mathcal{H}}$ requires both symmetries.
This difference, together with the difference in $(N_w-N_c)/2$,
distinguishes the two $\mathbb{Z}_2$ invariants deduced form the two models.

{\color{cyan}\indent{\em Discussion.}}---
The Josephson effect in Fig.~\ref{fig1}(b) was first proposed in ref.~\onlinecite{TRITSC}
and further studied in refs.~\onlinecite{Law} and~\onlinecite{Berg}, however, each one requires an {\it extra} symmetry.
With a mirror symmetry~\cite{TMSC}, a SC in class DIII decomposes into two insulators in class AIII
and a Majorana quartet at $\phi=\pi$ may be protected~\cite{TRITSC,Berg}.
With a mirror-gauge symmetry~\cite{MG}, the SC decouples into two SCs in class D
and each one may exhibit a fractional Josephson effect~\cite{Law}.
Indeed the TR symmetry is not essential for these cases,
as long as the mirror-like symmetries are unbroken.
Importantly, as first demonstrated in this Letter,
the fractional Josephson effect in Fig.~\ref{fig1}(b) can be protected by the TR symmetry even in the absence of any extra symmetry.

To discover the novel topological pumps summarized in Tables~\ref{tableone} and \ref{tabletwo},
we have provided the symmetry and homotopy arguments on ``bulk" invariants,
the stability analysis of ``boundary" consequences, and the Clifford algebra of representative models.
We are able to reproduce the well known $\mathbb{Z}_2$ invariant
for the $d_k=d_{\phi}=1$ case in class AII (D) where there is only TR (PH) symmetry.
Basically without the chiral symmetry one $\mathbb{Z}_2$ invariant associated with the winding number vanishes,
e.g., the one from Eq.~(\ref{4band}).
With these powerful approaches, our results can be generalized to another three symmetry classes with the chiral symmetry
and to the DIII cases with more than one anomalous parameters.
Besides the strong invariants that we have established,
we note there exists multiple weak invariants in lower dimensional subspaces of the extended BZ.

{\color{cyan}\indent{\em Acknowledgement.}}---
We thank A. Akhmerov, E. J. Mele, J. E. Moore, and Jianhui Zhou for helpful discussions.
FZ is supported by DARPA grant SPAWAR N66001-11-1-4110.
CLK is supported by a Simons Investigator award from the Simons Foundation.

\bibliographystyle{apsrev4-1}

\begin{thebibliography}{100}

\bibitem{RMP1}
 M. Z. Hasan and C. L. Kane, Rev. Mod. Phys. {\bf 82}, 3045 (2010).

\bibitem{Nature}
 J. E. Moore, Nature (London) {\bf 464}, 194 (2010).

\bibitem{RMP2}
 X. Qi and S. Zhang, Rev. Mod. Phys. {\bf 83}, 1057 (2011).

\bibitem{Ryu}
 A. P. Schnyder, S. Ryu, A. Furusaki, and A. W. W. Ludwig, Phys. Rev. B {\bf 78}, 195125 (2008).

\bibitem{Kitaev}
 A. Kitaev, AIP Conf. Proc. {\bf 1134}, 22 (2009).

\bibitem{Teo}
 J. C. Y. Teo and C. L. Kane, Phys. Rev. B {\bf 82}, 115120 (2010).

\bibitem{CN1}
 D. J. Thouless, M. Kohmoto, M. P. Nightingale, and M. den Nijs, Phys. Rev. Lett. {\bf 49}, 405 (1982).

\bibitem{CN2}
 J. E. Avron, R. Seiler, and B. Simon, Phys. Rev. Lett. {\bf 51}, 51 (1983).

\bibitem{Read}
 N. Read and D. Green, Phys. Rev. B {\bf 61}, 10267 (2000).

\bibitem{Kitaev-2001}
 A. Kitaev, Phys. Usp. {\bf 44}, 131 (2001);

\bibitem{Ivanov}
 D. A. Ivanov, Phys. Rev. Lett. {\bf 86}, 268 (2001).

\bibitem{Z2}
 C. L. Kane and E. J. Mele, Phys. Rev. Lett. {\bf 95}, 146802 (2005).

\bibitem{spin-pump}
 L. Fu, and C. L. Kane, Phys. Rev. B {\bf 74}, 195312 (2006).

\bibitem{Volovik}
 G. E. Volovik, The Universe in a Helium Droplet, Oxford University Press, (2003).

\bibitem{TMSC}
 F. Zhang, C. L. Kane, and E. J. Mele, Phys. Rev. Lett. {\bf 111}, 056403 (2013).

\bibitem{TRITSC}
 F. Zhang, C. L. Kane, and E. J. Mele, Phys. Rev. Lett. {\bf 111}, 056402 (2013).

\bibitem{WTI}
 For the $d_k-d_r=2$ case, we can use either a Rashba layer or a weak TI~\cite{TRITSC}.

\bibitem{Wong}
 C. L. M. Wong and K. T. Law, Phys. Rev. B {\bf 86}, 184516 (2012).

\bibitem{Nagaosa}
 S. Nakosai, J. C. Budich, Y. Tanaka, B. Trauzettel, and N. Nagaosa, Phys. Rev. Lett. {\bf 110}, 117002 (2013).

\bibitem{Berg}
 A. Keselman, L. Fu, A. Stern, and E. Berg, Phys. Rev. Lett. {\bf 111}, 116402 (2013).

\bibitem{Kane-D}
 L. Fu and C. L. Kane, Phys. Rev. B {\bf 79}, 161408(R) (2009).

\bibitem{Lutchyn-D}
 R. M. Lutchyn, J. D. Sau, S. Das Sarma, Phys. Rev. Lett. {\bf 105}, 077001 (2010).

\bibitem{Oreg-D}
 Y. Oreg, G. Refael, and F. von Oppen, Phys. Rev. Lett. {\bf 105}, 177002 (2010).

\bibitem{breaking-mirror}
This is a general feature of the Josephson effect for 1D topological SCs in class DIII.
For instance, Fig.3(b) in ref.~\onlinecite{TRITSC} becomes Fig.~\ref{fig1}(b) here, if we add a term which breaks mirror symmetry but preserves TR symmetry.

\bibitem{Nayak}
 C. Nayak, S. H. Simon, A. Stern, M. Freedman, and S. Das Sarma, Rev. Mod. Phys. {\bf 80}, 1083 (2008).

\bibitem{Fisher}
J. Alicea, Y. Oreg, G. Refael, F. von Oppen, and M. P. A. Fisher, Nature Phys. {\bf 7}, 412 (2011).

\bibitem{Moore-Balents}
 J. E. Moore and L. Balents, Phys. Rev. B {\bf 75}, 121306 (2007).

\bibitem{reduction}
 X.-L. Qi, T. Hughes, and S.-C. Zhang, Phys. Rev. B {\bf 78}, 195424 (2008).

\bibitem{boundary}
 F. Zhang, C. L. Kane, and E. J. Mele, Phys. Rev. B {\bf 86}, 081303(R) (2012).

\bibitem{Law}
 X.-J. Liu, C. L. M. Wong, and K. T. Law, arXiv:1304.3765 (2013).

\bibitem{MG}
 The normal state (order parameter) is even (odd) under the mirror symmetry.

%\bibitem{MKP}
% This is a pair of Majoranas that are TR partners and form a special fermion level.
% States with this level occupied and unoccupied are TR partners and thus TR symmetry acts like a supersymmetry that changes the fermion parity. More discussions are in ref.~\onlinecite{TRITSC,TMSC,Wong,Nagaosa}.

%%\bibitem{mirror}
%% Weak invariants are the strong invariants in lower dimensions. Invariants protected by a mirror symmetry~\cite{TMI,TMSC,Sato,Chiu,Furusaki}
%% are the invariants in a mirror invariant plane/line for each mirror eigenspace. Both cases simply reproduce those entries in the table.

\end{thebibliography}

\end{document}